# SpaceRaceEdu: desarrollo de un videojuego multijugador educativo para el autoestudio y la autoevaluación

Juan Jesús Roldán Gómez [a], Cristina Alonso Fernández [b] y Carlos Aguirre Maeso [c]

[a] Universidad Autónoma de Madrid, juan.roldan@uam.es, Departamento de Ingeniería Informática, Escuela Politécnica Superior, c/ Francisco Tomás y Valiente, nº11, 28049 Madrid, España

[b] Universidad Autónoma de Madrid, cristina.alonsof@uam.es, Departamento de Ingeniería Informática, Escuela Politécnica Superior, c/ Francisco Tomás y Valiente, nº11, 28049 Madrid, España

[c] Universidad Autónoma de Madrid, carlos.aguirre@uam.es, Departamento de Ingeniería Informática, Escuela Politécnica Superior, c/ Francisco Tomás y Valiente, nº11, 28049 Madrid, España

## Resumen

El proyecto de innovación docente "SpaceRaceEdu: desarrollo de un videojuego multijugador educativo para el autoestudio y la autoevaluación" se ha llevado a cabo bajo la convocatoria INNOVA de la Universidad Autónoma de Madrid durante el curso 2022-2023.

En este proyecto se ha desarrollado un prototipo funcional de SpaceRaceEdu: un videojuego multijugador con carácter social y educativo, que puede utilizarse tanto por parte del profesorado como actividad de formación y evaluación, como por parte del alumnado como herramienta para el estudio y evaluación.

SpaceRaceEdu enfrenta a varios equipos de estudiantes que tratan de lanzar un cohete antes que los demás. Para cumplir con este objetivo, deben reunir una serie de recursos recorriendo un escenario y respondiendo preguntas de diferentes tipos. Estas preguntas pueden ser introducidas por un profesor de acuerdo con los contenidos de su asignatura.

El videojuego busca un equilibrio entre competición y cooperación para favorecer la participación y el aprendizaje. La competición se produce entre los equipos que luchan por responder correctamente a todas sus preguntas antes que los rivales, mientras que la cooperación se produce entre los estudiantes del mismo equipo que pueden organizarse y apoyarse para ser más eficaces.

Palabras clave: Herramientas TIC, docencia, juegos serios, autoaprendizaje, autoevaluación.

## *SpaceRaceEdu: developing an educational multi-player videogame for self-study and assessment*

### *Abstract*

*The teaching innovation project "SpaceRaceEdu: development of an educational multiplayer video game for self-study and self-assessment" has been carried out under the INNOVA call of the Autonomous University of Madrid during the 2022-2023 academic year.*

*In this project, a functional prototype of SpaceRaceEdu has been developed: a multiplayer video game with a social and educational nature, which can be used both by teachers as a training and evaluation activity and by students as a tool for study and evaluation.*

*In SpaceRaceEdu, several student teams try to launch a rocket before everyone else. To meet this objective, they must gather a series of resources by going through a scenario and answering questions of different types. The teachers can introduce these questions according to the contents of their subject.*

*The videogame balances competition and cooperation to promote participation and learning. Competition occurs between teams who strive to answer all their questions correctly before their rivals. In contrast, cooperation occurs between students on the same team who can organize and support each other to be more effective.*

*Key words: ICT tools, teaching, serious games, self-learning, self-assessment.*

1. Introducción

La aplicación de las Tecnologías de la Información y la Comunicación (TIC) en la educación ha crecido sostenidamente en los últimos años. La irrupción de la pandemia de COVID-19 y las medidas para frenar su expansión, como los confinamientos o el trabajo remoto, provocaron un gran aumento en la necesidad y utilización de estas tecnologías a partir del año 2020. Herramientas de software que se empleaban esporádicamente o como soporte a las actividades presenciales, comenzaron a emplearse a diario y de forma masiva para realizar las clases y tutorías remotas [1].

La pandemia también revolucionó el ocio de millones de personas en todo el mundo. Las medidas para evitar la propagación del virus restringieron el ocio tradicional y produjeron el auge de alternativas digitales a través de internet: redes sociales, videojuegos y plataformas de streaming, tanto enfocadas hacia el cine y las series como a los videojuegos y contenidos variados.

Un caso relevante entre estas plataformas es Twitch, que experimentó un gran crecimiento en todos los sentidos: los canales se duplicaron, los espectadores concurrentes se triplicaron y las horas de emisión y visionado se duplicaron entre 2019 y 2021. Hubo un crecimiento de categorías no relacionadas con el juego, especialmente relevante en conversaciones (just chatting), ciencia y tecnología (science & technology), comida y bebida (food & drink) y música (music). También en categorías relacionadas con juegos centrados en la interacción social e interpretación de roles (role-playing) como Minecraft, Rust, Ark, Grand Theft Auto, Lost Ark y World of Warcraft [2].

Como se puede observar, algunos de los videojuegos más exitosos durante esas fechas tenían un importante carácter multijugador y social, como es el caso de Among Us. Este juego está ambientado en una nave espacial, donde los jugadores asumen roles de tripulantes o impostores. Los tripulantes deben completar tareas alrededor de la nave mientras intentan identificar a los impostores, quienes tienen la misión de sabotear las operaciones y eliminar a los tripulantes sin ser descubiertos. El juego combina los géneros de deducción social, crimen y misterio, rol y fiesta. A pesar de que surgió en 2018, alcanzó la popularidad en 2020, espoleado por la pandemia y las medidas de distanciamiento social [3].

Planteamos el proyecto de innovación docente "SpaceRaceEdu: desarrollo de un videojuego multijugador educativo para el autoestudio y la autoevaluación" para investigar la integración de estos dos recursos en la postpandemia: las herramientas TIC aplicadas al aprendizaje y los videojuegos multijugador de carácter social. Este proyecto se ha desarrollado en el marco de la convocatoria INNOVA de la Universidad Autónoma de Madrid del curso 2022-2023.

En las últimas dos décadas ha habido numerosas investigaciones en juegos serios con propósito educativo. El resultado han sido propuestas de juegos con diferentes géneros (aventuras, acción, simulación, puzle…), materias (lenguajes, programación, matemáticas…) y mecánicas (espacios, elección, habilidad…) [4]. La mayoría de estos juegos crean escenarios multijugador con elementos competitivos para fomentar la motivación de los jugadores [5], pero algunos también exploran la colaboración como medio para un aprendizaje más eficaz y divertido [6]. El proyecto ha buscado continuar con esta experiencia de los juegos serios para crear un juego multijugador de carácter social que combina competición y colaboración.

## 2. Objetivos

El objetivo principal del proyecto ha sido el desarrollo de un prototipo de SpaceRaceEdu, un videojuego multijugador con carácter social y educativo, que puede utilizarse tanto por parte del profesorado como actividad de formación y evaluación, como por parte del alumnado como herramienta para el estudio y evaluación.

Este objetivo se puede descomponer en los siguientes subobjetivos:

1. Recopilar opiniones de todos los agentes educativos (profesores y estudiantes) en todos los niveles formativos (primaria, secundaria y universidad).
2. Diseñar el videojuego: reglas del juego, arte de escenarios y personajes, control de personajes, comunicación, etc.
3. Desarrollar un primer prototipo que permita llevar a cabo una partida de un solo jugador.
4. Desarrollar un segundo prototipo con funciones de red que habiliten una partida multijugador.
5. Desarrollar una interfaz de administrador que permita crear y gestionar una partida.
6. Recopilar opiniones de todos los agentes educativos en todos los niveles formativos.

## 3. Diseño y desarrollo

### *3.1. Objetivo del juego*

SpaceRaceEdu sitúa a los jugadores en un centro de investigación en medio de una carrera espacial. En este entorno varios equipos compuestos por varios jugadores deben competir para lanzar su cohete en primer lugar. Para efectuar este lanzamiento, los jugadores deben reunir una cantidad suficiente de energía recorriendo el escenario, interactuando con ciertos elementos y respondiendo con acierto a las preguntas.

### *3.2. Fundamentos educativos*

Desde el punto de vista educativo, el videojuego se ha diseñado siguiendo dos principios: crear un entorno seguro y divertido para el autoestudio y la autoevaluación y buscar un equilibrio entre la competición y la colaboración.

Respecto al primer principio, se buscó una diferenciación entre el juego propuesto para el autoestudio y la autoevaluación y los entornos de estudio y evaluación habituales. Esto es necesario en los dos casos de uso considerados: por un lado, permite sacar el máximo partido al juego como actividad de clase propuesta por el profesorado y, por el otro, fomenta su uso como recurso del alumnado fuera del horario de clase. Para recrear este entorno seguro se recompensan los aciertos, pero no se penalizan los fallos, sino que se permite repetir la pregunta en un cierto tiempo hasta que se acierte.

Respecto al segundo principio, se consideró importante alcanzar un balance entre la competición y la colaboración para potenciar sus ventajas y compensar sus potenciales inconvenientes. El factor competitivo busca mejorar la atención e implicación de los estudiantes en el desarrollo de la partida proporcionándoles motivación y entretenimiento. Por su parte, el factor colaborativo busca fomentar el desarrollo de competencias trasversales como el trabajo en equipo en un entorno en que es claramente beneficioso.

### 3.3. Requisitos técnicos

A nivel técnico se consideraron los siguientes requisitos para el videojuego:

1. Escenario: Un centro de investigación espacial.
2. Jugadores: Hasta cuatro equipos de hasta diez jugadores.
3. Interfaces: Una para los jugadores y otra para el administrador.
4. Interfaz de jugador: Permite buscar una partida, unirse a ella, elegir un equipo y jugar la partida.
5. Interfaz de administrador: Permite crear una partida, generar preguntas, cargar una bolsa de preguntas, supervisar el estado de la partida y finalizar la partida.
6. Control de jugadores: Teclas de flechas para moverse por el escenario y botón izquierdo del ratón para interactuar con objetos.
7. Preguntas: Cuatro tipos de preguntas; respuesta múltiple, respuesta numérica, ordenación y clasificación.
8. Comunicación: Externa al videojuego; directa en partidas presenciales y aplicación de llamada en partidas remotas.

### 3.4. Partida del juego

La partida es creada por el administrador, que puede ser un profesor o un estudiante según si se está jugando en clase o fuera de ella, seleccionando "host" en el menú principal de la Figura 1. A partir de ahí, podrá cargar un banco de preguntas existente o crear uno nuevo usando las vistas de creación de banco y de pregunta de la Figura 1. Este administrador podrá supervisar el curso de la partida y acceder a un informe tras su finalización con la vista de supervisión de partida de la Figura 1.

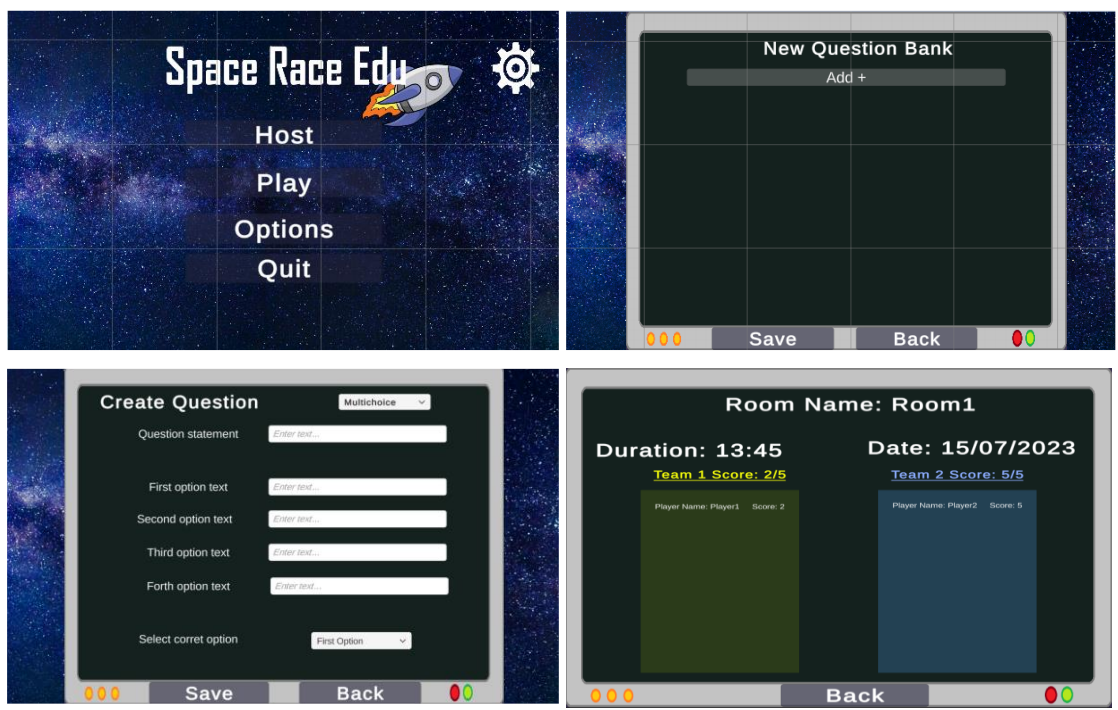

*Figura 1. Menú principal (arriba izquierda), creación de banco de preguntas (arriba derecha), creación de pregunta (abajo izquierda) y supervisión de partida (abajo derecha).*

Por su parte, los jugadores deben seleccionar "play" en el menú principal de la Figura 1 para poder unirse a la partida creada por el administrador y seleccionar el equipo con el que quieren jugar. Al acceder a la partida les aparecerá la lista de tareas a realizar como se muestra en la Figura 2. Esta lista también aparecerá cada vez que completen una tarea para recordarles las que restan para finalizar la partida. Los jugadores controlarán unos personajes humanoides por un escenario con ambientación tecnológica como se muestra en la Figura 2, interactuando con diferentes máquinas para acceder a las preguntas que les permiten avanzar en el juego.

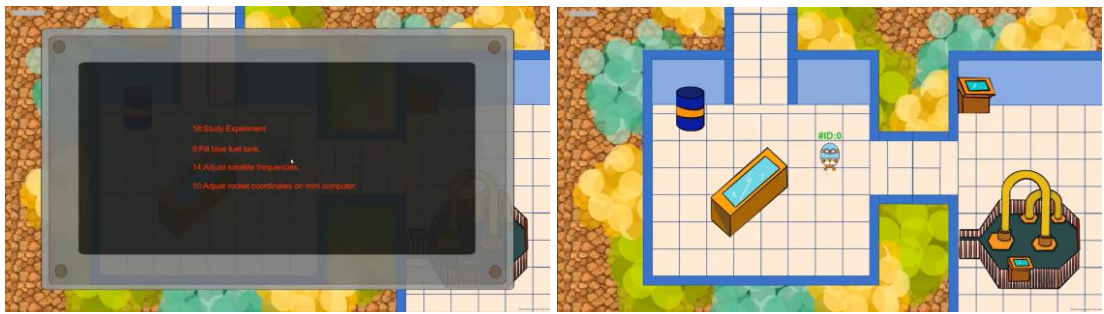

*Figura 2. Lista de tareas por hacer (izquierda) y personaje en el escenario (derecha).*

Las preguntas pueden ser de cuatro tipos como se puede observar en la Figura 3: selección múltiple, donde el jugador tiene que elegir una o varias respuestas correctas; numérica, donde el jugador debe responder a la pregunta introduciendo un número; ordenación, donde el jugador debe ordenar cuatro respuestas atendiendo al criterio indicado; y clasificación, donde el jugador debe clasificar cuatro respuestas en dos categorías.

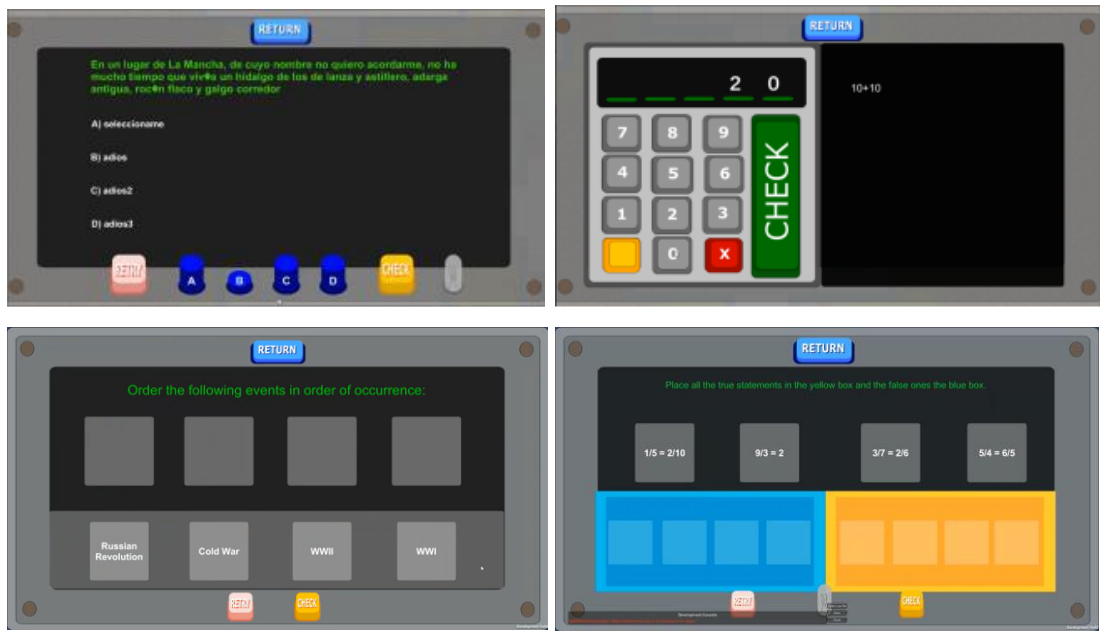

*Figura 3. Preguntas de respuesta múltiple (arriba izquierda), respuesta numérica (arriba derecha), ordenación de respuestas (abajo izquierda) y clasificación de respuestas (abajo derecha).*

La partida acaba cuando un equipo ha conseguido realizar todas las tareas o, lo que es lo mismo, responder correctamente a todas las preguntas asignadas por el administrador.

## 4. Pruebas y resultados

El prototipo de SpaceRaceEdu se ha sometido a dos tipos de pruebas: técnicas y educativas.

Las primeras han servido para validar el cumplimiento de los requisitos de diseño del videojuego, establecidos tras recopilar opiniones de profesores y estudiantes en primaria, secundaria y universidad, realizar una revisión del estado del arte de juegos serios centrados en la educación e incorporar nuestras propias ideas sobre la mecánica del juego. Además, estas pruebas han servido para detectar errores que sólo se manifiestan durante la ejecución del programa, sobre todo cuando se lleva a cabo una partida en red con varios jugadores en diferentes localizaciones.

Las segundas han servido para recoger las opiniones de los agentes educativos, permitiendo conocer el potencial y los límites del videojuego. En concreto, se grabaron varios vídeos del juego y se enviaron junto con un cuestionario a profesores y estudiantes de primaria, secundaria y universidad. Los estudiantes valoraron la mecánica del juego con un 3.9 y la comodidad de jugar en clase con un 4.4 sobre 5. Además, más de dos tercios se mostraron partidarios de utilizarlo como recurso para estudiar en casa. Por su parte, un 36% de los profesores lo vieron aplicable totalmente y un 55% parcialmente en sus asignaturas. El 64% lo vieron como una herramienta útil para animar a sus estudiantes a trabajar en equipo. Sólo una minoría del 18% lo vio aplicable para actividades de evaluación, pero una mayoría del 90% lo vio útil para el estudio fuera del aula.

## 5. Conclusiones y trabajos futuros

Este proyecto ha dado lugar a un prototipo funcional de SpaceRaceEdu con interfaces de jugador y administrador y modos para uno y varios jugadores. Este prototipo cumple con los requisitos establecidos por los investigadores tras recoger opiniones de los agentes educativos, estudiar los precedentes de juegos serios y plantear su propio diseño para el juego. En las pruebas, el administrador ha podido crear una partida, cargar una bolsa de preguntas y supervisar su desarrollo, mientras que los jugadores han podido desplazarse por el escenario, interactuar con los objetos, responder las preguntas y completar la partida. Por último, se han facilitado un conjunto de vídeos del juego a profesores de primaria, secundaria y universidad, que han dado una realimentación muy valiosa para trabajar en la siguiente versión del juego.

En un proyecto futuro se espera probar el videojuego en condiciones de aprendizaje reales y variadas. Para ello se realizarán dos experimentos: en el primero, los profesores propondrán y dirigirán una partida del juego en horario de clase como una actividad práctica que complemente sus clases teóricas, mientras que, en el segundo, los estudiantes podrán crear y jugar partidas a su interés fuera del horario de clase para preparar las evaluaciones de las asignaturas. Con objeto de obtener resultados más amplios y determinar el público idóneo para el juego, estas pruebas se llevarán a cabo en los niveles de primaria, secundaria y universidad. De esta manera se espera detectar y corregir posibles errores, obtener realimentación más detallada de profesores y estudiantes, encontrar el ámbito de aplicación idóneo para el juego y valorar su impacto en el proceso educativo.

Para más información sobre el diseño, desarrollo e implementación del videojuego SpaceRaceEdu se pueden consultar las referencias [7-10]

## Agradecimientos

## Referencias

[1] Ozdamli, F., & Karagozlu, D. (2022). Online education during the pandemic: A systematic literature review. International Journal of Emerging Technologies in Learning (iJET), 17(16), 167-193.

[2] Scerbakov, A., Pirker, J., & Kappe, F. (2022, January). When a Pandemic Enters the Game: The Initial and Prolonged Impact of the COVID-19 Pandemic on Live-Stream Broadcasters on Twitch. In HICSS (pp. 1-10).

[3] Bump, A., & Şengün, S. (2021). Among Us and Its Popularity During COVID-19 Pandemic. In Encyclopedia of Computer Graphics and Games (pp. 1-4). Springer, Cham.

[4] Wang, C., & Huang, L. (2021). A Systematic Review of Serious Games for Collaborative Learning: Theoretical Framework, Game Mechanic and Efficiency Assessment. International Journal of Emerging Technologies in Learning, 16(6).

[5] Bawa, P., Watson, S. L., & Watson, W. (2018). Motivation is a game: Massively multiplayer online games as agents of motivation in higher education. Computers & Education, 123, 174-194.

[6] Wendel, V., Gutjahr, M., Göbel, S., & Steinmetz, R. (2013). Designing collaborative multiplayer serious games: Escape from Wilson Island—A multiplayer 3D serious game for collaborative learning in teams. Education and Information Technologies, 18, 287-308.

[7] Hadeed, L. (2021). SpaceRaceEdu: Development of a Gamified Educational Tool for Self-Study and Evaluation. Trabajo Final de Grado, Escuela Politécnica Superior, Universidad Autónoma de Madrid.

[8] Martínez, J. (2022). SpaceRaceEdu: Desarrollo de arquitectura y back-end para videojuego multijugador educativo. Trabajo Final de Grado, Escuela Politécnica Superior, Universidad Autónoma de Madrid.

[9] Maza, C. (2023). SpaceRaceEdu: Implementación de modo multi-jugador para videojuego educativo. Trabajo Final de Grado, Escuela Politécnica Superior, Universidad Autónoma de Madrid.

[10] Mérida, J.C. (2023). SpaceRaceEdu: Interfaz de profesor para videojuego educativo multi-jugador. Trabajo Final de Grado, Escuela Politécnica Superior, Universidad Autónoma de Madrid.